\documentclass[]{iopart}
\usepackage{graphicx}
\begin{document}
\bibliographystyle{iopart-num}
\title{Evidence of delocalized excitons in amorphous solids}
\author{Fabrizio Messina, Eleonora Vella, Marco Cannas and Roberto Boscaino}
\ead{eleonora.vella@fisica.unipa.it}
\address{Dipartimento di Scienze Fisiche ed Astronomiche,
Universit$\grave{a}$ di Palermo, Via Archirafi 36, I-90123 Palermo,
Italy}
\begin{abstract}
We studied the temperature dependence of the absorption
coefficient of amorphous SiO$_2$ in the range from 8 to 17.5~eV
obtained by Kramers-Kronig dispersion analysis of reflectivity
spectra. We demonstrate the main excitonic resonance at 10.4~eV to
feature a close Lorentzian shape red-shifting with increasing
temperature. This provides a strong evidence of excitons being
delocalized notwithstanding the structural disorder intrinsic to
the amorphous system. Excitons turn out to be coupled to an
average phonon mode of 83~meV energy.
\end{abstract}
 \maketitle Excitons are neutral
quasi-particles consisting in an electron-hole pair bound via
Coulomb interaction. One of the most important features of
excitons is their mobility, providing a fundamental mechanism of
site-to-site energy transfer in condensed matter without charge
transport. While being one of the more complex manifestations of
the electronic band structure of a solid, excitons play a key role
in many processes occurring not only in solid state but also in
nanoscale systems as well as in molecular materials. These include
light emission in semiconductor devices,  energy capture in
photovoltaic systems \cite{KoizumiScience01,TaniyasuNature06,ScholesNatMat06} or generation of
point defects upon exposure to high power laser light
\cite{ShlugerPRB00,FukataAPL03,BeigiPRL05}. Furthermore, exciton dynamics may feature
collective phenomena such as Bose-Einstein condensation which are
currently the subject of intense research \cite{DengScience02,KasprzakNature06}.
Exciton mobility in a crystal is influenced by their interaction
with phonons, leading in some cases to the trapping of the
quasi-particle. In disordered systems lacking translational order
such as glasses, the very existence of excitons was disputed for
some time, but it is widely accepted now. However, the properties
of excitons in such systems are still debated \cite{VlamingPRB09,HuijserJAmChemSoc08},
those concerning the influence of structural disorder on their
mobility being poorly understood as yet. This is particularly true
for wide band-gap insulators, of which amorphous silicon dioxide
(a-SiO$_2$) is an archetypal system, because their excitonic
absorption is difficult to investigate experimentally requiring
vacuum-UV (VUV, $\lambda<$200~nm) optical technologies. a-SiO$_2$
is a solid of primary technologic importance in optics and
microelectronics. It has lately received renewed attention aiming
to further improve its optical properties in view of prospective
VUV applications \cite{Erice,SkujaSPIE03}. A full comprehension of
exciton properties needs the thorough understanding of the
fundamental band-to-band transition, which is among the basic open
problems in the physics of a-SiO$_2$ \cite{Erice,VellaPRB09}. This
study has often been aided by the comparison with $\alpha$-quartz
(c-SiO$_2$), the most common crystalline polymorph of silicon
dioxide. Several experimental results, such as the detection of
photoconductivity starting from $\sim$9~eV \cite{WeinbergPRB79,DiStefanoSSC71},
led to estimate the latter value as the band-gap of a- and
c-SiO$_2$
\cite{WeinbergPRB79,DiStefanoSSC71,PlatzoderPSS68,ItohPRB89,BosioEPL93}. Some
studies questioned these findings, by reporting a marked increase
of photoconductivity and photoemission signals only above 11~eV,
accompanied by qualitative variations of the photoelectron energy
distribution \cite{TrukhinJNCS92,Erice}. For this reason the
actual band-gap of SiO$_2$ was proposed to lay around 11~eV
\cite{TrukhinJNCS92}, but this is still a matter of debate. The
experimental investigation of these issues in SiO$_2$ requires the
combined use of different techniques. Optical absorption (OA)
spectroscopy can be applied from $\sim$8 to $\sim$9~eV, where the
absorption coefficient $\alpha$ is smaller than
10$^3$-10$^4$~cm$^{-1}$ and both a- and c-SiO$_2$ feature the
exponential absorption region usually referred to as the Urbach
tail \cite{VellaPRB09}. Reflectance spectroscopy is necessary at
higher energies, where $\alpha$ becomes exceedingly high for OA so
that absorption can be only indirectly inferred by Kramers-Kronig
(KK) analysis of reflectivity data. KK analysis can be quite
tricky due to the need of reliable measurements over an extended
energy range. This has limited the number of KK investigations on
SiO$_2$ to date \cite{PhilippJPCS71,BosioEPL93,TanPRB05} and
consequently hindered the comprehensive understanding of near-edge
absorption in wide band-gap amorphous systems. The lowest energy
peak in the absorption spectra of a- and c-SiO$_2$ is at 10.4~eV
\cite{PlatzoderPSS68,Erice,TanPRB05,PhilippJPCS71} and is widely
accepted to be an excitonic transition, based on its temperature
dependence observed in c-SiO$_2$ from 300 to 600~K
\cite{PlatzoderPSS68,PhilippJPCS71}. However, almost no
experimental data exist on the temperature dependence of the
10.4~eV peak below 300~K, especially in a-SiO$_2$. Most important,
the shape of this resonance is currently an open issue. Well
established theoretical models describe the broadening of an
exciton resonance in a crystal as resulting from exciton-phonon
coupling (EPC) and the excitonic lineshape as arising from the
competition between mobility of excitons and thermal fluctuations
of their energy \cite{SchreiberJPSJ82,Toyozawa}. Although no models exist
for amorphous solids, we argue that a similar description could be
used for them as well since it was suggested that the effects of a
small structural disorder in crystalline systems are
indistinguishable from those of thermal fluctuations
\cite{SchreiberJPSJ82,Toyozawa}. In this Letter we contribute to the
clarification of these topics by providing experimental data on
the temperature dependence of the reflectivity spectra of the
model system a-SiO$_2$ below 300~K. Our data yield strong evidence
of the existence of delocalized, closely crystal-like, excitons in
an amorphous system.

We measured reflectivity $R(E)$ for 9~eV$<$$E$$<$21~eV (wavelength
accuracy $\pm$0.1~nm) of commercial synthetic a-SiO$_2$ and z-cut
c-SiO$_2$ samples, featuring industrial grade optical polishing.
Measurements were carried out in ultra-high vacuum in near-normal
geometry under synchrotron radiation at beamline I in HASYLAB
(DESY). The reflected beam was detected by a photomultiplier
coated by C$_7$H$_5$NaO$_3$. Each acquired spectrum was scaled for
the spectral density curve of the excitation beam. Experimental
spectra were extended on the low and high energy tails using
literature data down to 0~eV \cite{PhilippJPCS71} and the ordinary
analytical form in the free-electron gas approximation
\cite{JonesJMaterRes99}, $R(E)\propto E^{-4}$, for $E>$21~eV. Finally,
standard KK analysis was performed on $R(E)$ in order to obtain
the absorption coefficient $\alpha(E)$. The influence on the
10.4~eV peak of different reasonable choices for the high energy
extrapolation was verified to be very poor.
\begin{figure}
\begin{center}
\includegraphics[width=8.5cm]{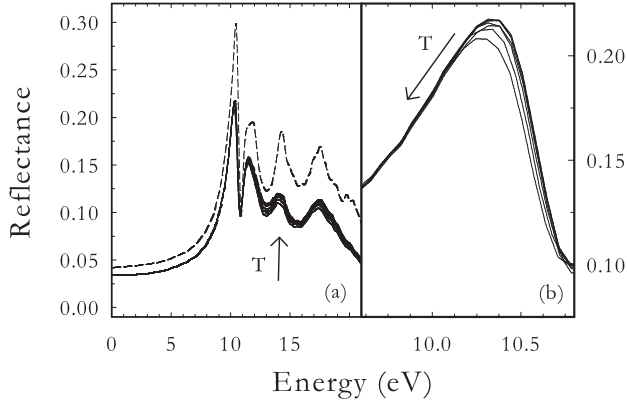}
\caption{\label{SpettriRifl}(a) Reflectivity of a-SiO$_2$ from 10
to 300~K by steps of 50~K (full lines), and reflectivity of
c-SiO$_2$ at 10~K (dashed line). (b) Enlargement of the peak at
$\sim$10.3~eV in a-SiO$_2$. The arrows show the direction of
growing temperatures.}
\end{center}
\end{figure}
Figure~\ref{SpettriRifl}-a shows the reflectivity spectra of
a-SiO$_2$ from 10 to 300~K and that of c-SiO$_2$ at 10~K for
comparison. In a-SiO$_2$ (c-SiO$_2$) spectra feature four main
peaks: a first narrow and more intense peak at $\sim$10.3~eV
($\sim$10.4~eV) and three other broader peaks at $\sim$11.5
($\sim$11.4~eV), $\sim$14.0 ($\sim$14.1~eV) and $\sim$17.3~eV
($\sim$17.3~eV) respectively. The positions of the peaks in a- and
c-SiO$_2$ agree qualitatively with those reported in literature
\cite{PhilippJPCS71,BosioEPL93,TanPRB05,Erice}. In a-SiO$_2$ the
peaks at $\sim$10.3 and $\sim$11.5~eV change in both their
positions and amplitudes as a function of temperature, whereas the
peaks at $\sim$14.0 and $\sim$17.3~eV increase their intensities
without noticeably changing their shapes with increasing
temperature. As visible in Fig.~\ref{SpettriRifl}-b, with
increasing temperature the peak at $\sim$10.3~eV red-shifts while
its amplitude decreases.
\begin{figure}
\begin{center}
\includegraphics[width=8.5cm]{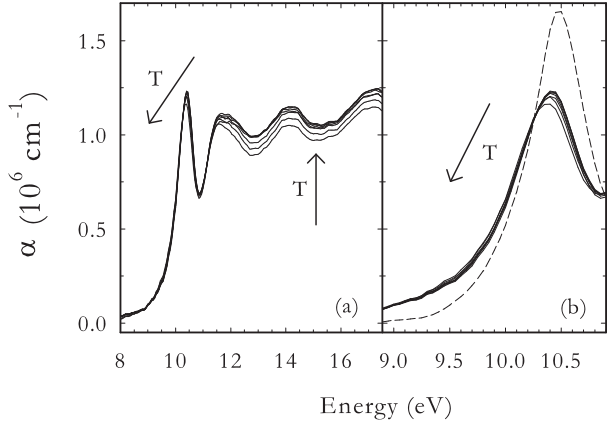}
\caption{\label{SpettriAlfa}(a) Absorption coefficient of
a-SiO$_2$ from 10 to 300~K by steps of 50~K. (b) Enlargement of
the peak at $\sim$10.4~eV in a-SiO$_2$ (full lines) and in
c-SiO$_2$ (dashed line). The arrows show the direction of growing
temperatures.}
\end{center}
\end{figure}
In Fig.~\ref{SpettriAlfa}-a the absorption coefficient of
a-SiO$_2$ is presented as calculated by KK analysis on spectra in
Fig.~\ref{SpettriRifl}. Four main peaks are present as well: at
$\sim$10.4, at $\sim$11.6, at $\sim$14.1 and at $\sim$17.3~eV.
Figure~\ref{SpettriAlfa}-b shows that the peak at $\sim$10.4~eV
red-shifts with increasing temperature while its amplitude
slightly decreases, moreover it is evident that in c-SiO$_2$ the
peak is narrower and more intense. Since the analysis of the peak
at $\sim$10.4~eV is the main purpose of the present work, the
properties of the other peaks will not be discussed in the
following.
\begin{figure}
\begin{center}
\includegraphics[width=8.5cm]{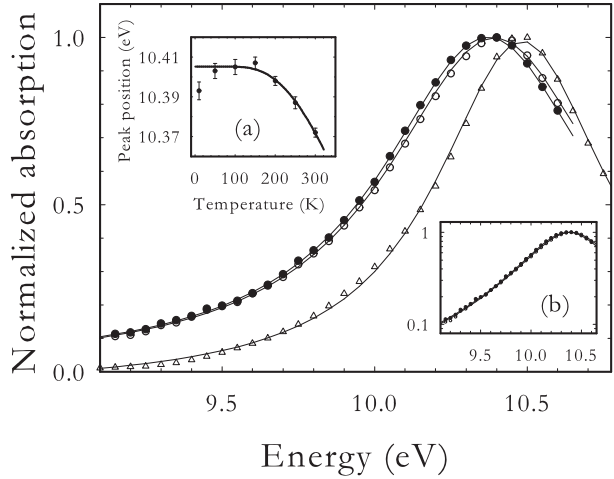}
\caption{\label{PrimoPiccoFitLoren} Normalized absorption at 10~K
(open circles) and at 300~K in a-SiO$_2$ (filled circles), at 10~K
in c-SiO$_2$ (triangles). Continuous lines were obtained via a
Lorentzian fit of each spectrum from 9 to 10.6~eV. (a) Position of
the peak in a-SiO$_2$ as a function of temperature (b)
Semilogarithmic plot of the same data on a-SiO$_2$ as in the main
panel.}
\end{center}
\end{figure}
The peak at $\sim$10.4~eV turns out to have a very good Lorentzian
profile at all temperatures:
\begin{equation}
\alpha(E,T)=\alpha_0\left[1+\left(\frac{E-(E_0^0+\Delta_0(T))}{\Gamma_0(T)/2}\right)^2\right]^{-1}
\label{LorentzianProfile}
\end{equation}
Figure~\ref{PrimoPiccoFitLoren} presents the normalized
$\sim$10.4~eV peak of a-SiO$_2$ at 10 and 300~K and that of
c-SiO$_2$ at 10~K, together with Lorentzian best fit curves:
agreement with experimental data is very close even on the left
band wings, as visible from the semi-logarithmic plot in inset
(b). We verified that introducing a baseline accounting for
band-to-band absorption does not alter appreciably the Lorentzian
lineshape, only slightly decreasing the width. The position of the
peak (inset (a)), as obtained by the Lorentzian fitting, is
approximately constant around a value of $\sim$10.41~eV from 10 to
150~K, whereas it shifts to lower energies for higher
temperatures. No change of the full width at half maximum (FWHM)
of the peak is appreciable within the experimental error: its
value is $(0.85\pm0.08)$~eV. In c-SiO$_2$ at 10~K the peak is at
10.5~eV and its FWHM is $(0.69\pm0.04)$~eV.

Although present data do not appear in contrast with previous ones
\cite{TanPRB05,Erice,PhilippJPCS71}, the accuracy of our
measurements allows an unprecedented characterization of the
$\sim$10.4~eV peak, leading to unanticipated insights on the
physics of excitons in a-SiO$_2$. We demonstrated the excitonic
peak at $\sim$10.4~eV to have a remarkably good Lorentzian
lineshape at all the examined temperatures. The absorption
lineshape of excitons in a crystal is basically governed by the
competition between their mobile nature and their tendency to
localization in the lattice due to EPC. The former is measured by
the width $B$ of the energy band of excitons, proportional to
site-to-site transfer rate, while the latter is measured by the
root mean square amplitude $D$ of the fluctuations of exciton
energy due to the thermal activated vibrations of the lattice.
Very general theoretical arguments (Kubo oscillator) \cite{KuboJMathPhys63}
strongly suggest that Lorentzian absorption lineshapes are
expected when mobility effects ($B$) prevail over energy
fluctuations ($D$). Indeed, the broadening of a resonance whose
proper energy undergoes random Gaussian fluctuations of amplitude
$D$ can be shown to be either Lorentzian or Gaussian when $D$ is
respectively much smaller or larger than $\hbar/\tau$, where
$\tau$ is the correlation time of the fluctuations. We physically
expect the mean time $\hbar/B$ between site-to-site hops of
excitons to basically represent $\tau$ (neglecting inter-site
correlations). Thus, the broadening is Lorentzian if $D$$\ll$$B$.
A detailed model of exciton-phonon interaction in solids
\cite{SchreiberJPSJ82,Toyozawa} was developed by Toyozawa, confirming the
same conclusion: if the ratio $D/B$ $\ll$1 and the mobile nature
of the exciton is thus prevailing (weak scattering limit), the
exciton absorption lineshape is Lorentzian. This can occur either
if EPC is inefficient (small $D$) or if excitons are mobile enough
(large $B$) to compensate for its effects. In this case, in
Eq.~\ref{LorentzianProfile} $E_0^0$ is the position of the
unperturbed exciton line, while $\Delta_0(T)$ and $\Gamma_0(T)$
are the real and imaginary parts of the self-energy of the exciton
in the phonon field, giving rise to a temperature dependent shift
($\Delta_0(T)$) and broadening ($\Gamma_0(T)$) of the exciton
line. The FWHM=2$\Gamma_0(T)$ is given by $cD^2/B$, where the
predicted value of the constant $c$ depends on the detailed
features of the exciton band \footnote{if multi-phonon scatterings
are taken into account, $c$$\sim$$6.3\pi$ \cite{SchreiberJPSJ82,Toyozawa}}.
The effect of increasing structural disorder (e.g. dislocations,
vacancies, interstitial atoms or impurities) can be argued to be
qualitatively the same as that of increasing temperature, in that
both lead to an increase of $D$. For thermal or structural
disorder so large or for $B$ so small that $D/B$ $\gg$1 (strong
scattering limit), the model foresees the overall exciton
lineshape to be Gaussian. While Lorentzian exciton lineshapes have
been reported in literature for several crystalline systems at low
temperatures \cite{SanoJPhysSocJapan69,PiacentiniSSC1975}, our analysis shows that Toyozawa's model
can be extended to an amorphous system. Present data clearly point
out that from 10 to 300~K excitons mobility is so large as
compared to the effects of both static disorder and EPC in
a-SiO$_2$ that the exciton band keeps its Lorentzian profile, not
presenting any tendencies towards a Gaussian one even at the
highest considered temperature (300~K). This result is remarkable
because exciton mobility is ultimately related to the
translational symmetry of the lattice. Thus, the lack of long
range order typical of an amorphous solid turns out to have a
negligible effect on the properties of excitons, meaning that the
effect of static disorder in a-SiO$_2$ is low enough to closely
preserve crystal-like delocalization of excitons. Surprisingly
enough, it appears that, at least at low temperature, the
difference between the excitonic properties of a- and c-SiO$_2$
comes down to a greater width of the first exciton absorption peak
in the amorphous system, $(0.85\pm0.08)$~eV instead than
$(0.69\pm0.04)$~eV, likely due to structural disorder introducing
small additional site-to-site fluctuations of the exciton energy
contributing to $D$. Furthermore, present results are completely
inconsistent with the model widely accepted in SiO$_2$ literature
claiming a strong localization and consequently a low mobility of
excitons in a-SiO$_2$ \cite{TrukhinJNCS92,Erice}. The evidence
that the peak is accurately described by a simple Lorenztian
function seriously casts doubt on its previously proposed
decomposition into at least four Gaussian sub-bands ascribed to
several types of localized excitons \cite{TrukhinJNCS92,Erice}.
Finally we note that the behavior of SiO$_2$ is deeply different
from that of other systems such as Si, where a dramatic change of
the shape and broadening of the first absorption peak are observed
when going from the crystal to the amorphous solid
\cite{AspnesPRB84}.

The value of $B$ can be approximately estimated as the separation
in energy between the first excitonic peak and the region where
the Urbach tail (8.0-8.3~eV) is observed in both a- and c-SiO$_2$
\cite{VellaPRB09}: i.e. $B$$\sim$2~eV. From this value and
comparing the above mentioned theoretical expression for the FWHM
with our data, we estimate $D_a$$\sim$0.30~eV in a-SiO$_2$ and
$D_c$$\sim$0.27~eV in c-SiO$_2$. Fluctuations of the exciton
energy of such an order of magnitude imply a rather efficient
coupling of excitons with vibrational modes, being the weak
scattering condition anyway ensured by the large extension of the
excitonic band, while the influence of structural disorder
($D_a$$\sim$$D_c$) is very poor, as anticipated. Thus, SiO$_2$
turns out to belong to that class of materials referred to as
featuring a strong EPC but a weak scattering
\cite{SchreiberJPSJ82,Toyozawa}. A similar behavior can be observed in some
very wide band-gap crystals such as LiF and NaF, having relatively
wide (FWHM$\sim$0.4~eV in LiF) yet Lorentzian exciton lineshapes
\cite{SanoJPhysSocJapan69,PiacentiniSSC1975} and, as SiO$_2$, all giving rise to self-trapped
exciton luminescence. Toyozawa's model predicts the position of
the excitonic peak to depend on temperature according to:
$E_0(T)=E_0^0+\Delta_0^0\coth\left(\hbar\omega_0/2k_BT\right)$,
where the parameter $\Delta_0^0$ represents the effect of the
zero-point vibrations on the shift of the peak and $\hbar\omega_0$
is the energy of thermal vibrations described using Einstein's
model. The effect of possible structural disorder is included in
$E_0^0$. As evident in Fig.\ref{PrimoPiccoFitLoren}-(a), this
equation well describes the observed trend with
$E_0^0$=$(10.79\pm0.05)$~eV, $\Delta_0^0$=$(-0.38\pm0.05)$~eV and
$\hbar\omega_0$=$(0.083\pm0.002)$~eV, with the only possible
exception of the point at 10~K \footnote{Admittedly present
results do not allow to establish if the observed scatter of this
datum points out a low temperature deviation from Toyozawa's
theory and, thus, it was not included in the fitting procedure.}.
Since $E_0^0$ represents the position of the first exciton
absorption line in the absence of interaction with phonons, our
data show that the band-gap, or mobility edge, of a-SiO$_2$ must
be located at energies higher than 10.79~eV. Finally, according to
Toyozawa's theory, the width should have the same temperature
dependence as the peak position, i.e. governed by
$\coth(\hbar\omega_0/2k_BT)$. Under this assumption and using for
$\hbar\omega_0$ the value reported above, the expected variation
of the width from 10 to 300~K is 8$\%$, corresponding to 0.07~eV,
and thus not detectable within our 0.08~eV uncertainty on the
FWHM. On the other side, we believe the observed decrease of the
amplitude of the peak (Fig.~\ref{SpettriAlfa}), measurable with a
better accuracy than the width, to be an indirect evidence of the
increase of the width of the line itself. In conclusion, the
observed behavior of above-edge absorption as obtained by KK
dispersion analysis of the reflectance spectra of a-SiO$_2$ from
10 to 300~K demonstrates the applicability of Toyozawa's theory
for excitonic lineshapes to an amorphous system. The remarkably
good and temperature independent Lorentzian profile of the main
excitonic peak at $\sim$10.4~eV proves that excitons in a-SiO$_2$
are highly mobile and retain their delocalized nature, albeit both
the disorder peculiar of the amorphous structure and a strong EPC.
The average vibrational energy of phonons interacting with
excitons ($\hbar\omega_0$=83~meV) and a 10.79~eV lower threshold
for a-SiO$_2$ band-gap were determined. We acknowledge grant
received from DESY and support from the LAMP group.
\section*{References}

\providecommand{\newblock}{}

\end{document}